\documentclass{jaa}

\usepackage{graphicx}
\usepackage{amsmath}
\usepackage{xcolor}
\usepackage{natbib}
\usepackage{footmisc}
\usepackage{rotating,longtable}
\usepackage{lineno}

\usepackage[T1]{fontenc}
\usepackage{ae,aecompl}

\def\apjl{ApJ}
\def\apjs{ApJS}
\def\aap{A\&A}
\begin{document}\sloppy

\title{Teutsch 76: a Deep Near-Infrared Study}


\author{Saurabh Sharma\textsuperscript{1,*}, Lokesh Dewangan\textsuperscript{2},  Neelam Panwar\textsuperscript{1},  Harmeen Kaur\textsuperscript{3},  Devendra K. Ojha\textsuperscript{4},  Ramkesh Yadav\textsuperscript{5}, Aayushi Verma\textsuperscript{1},  Tapas Baug\textsuperscript{6},  Tirthendu Sinha\textsuperscript{6},   Rakesh Pandey\textsuperscript{2}, Arpan Ghosh\textsuperscript{1}, and Tarak Chand\textsuperscript{1}}

\affilOne{\textsuperscript{1}Aryabhatta Research Institute of Observational Sciences (ARIES), Manora Peak, Nainital, 263 001, India\\}
\affilTwo{\textsuperscript{2}Astronomy and Astrophysics Division, Physical Research Laboratory (PRL), Navrangpura, Ahmedabad - 380009, India\\}
\affilThree{\textsuperscript{3}Center of Advanced Study, Department of Physics DSB Campus, Kumaun University Nainital, 263002, India\\}
\affilFour{\textsuperscript{4}Tata Institute of Fundamental Research (TIFR), Homi Bhabha Road, Colaba, Mumbai - 400 005, India\\}
\affilFive{\textsuperscript{5}National Astronomical Research Institute of Thailand (NARIT), Chiang Mai 50200, Thailand\\}
\affilSix{\textsuperscript{6}Satyendra Nath Bose National Centre for Basic Sciences (SNBNCBS), Block-JD, Sector-III, Salt Lake, Kolkata-700 106, India\\}


\twocolumn[{

\maketitle

\corres{saurabh@aries.res.in}

\msinfo{15 November 2022}{15 November 2022}

\begin{abstract}
        We have performed a detailed analysis on the Teutsch 76 (T76) open cluster using the deep near-infrared (NIR) observations taken with
	the TANSPEC instrument mounted on the 3.6m Devasthal Optical Telescope (DOT)
        along with the recently available high quality proper motion data from the {\it Gaia} data release 3 and deep photometric data from Pan-STARRS1 survey.
        We have found that the T76 cluster is having a central density concentration with circular morphology,
        probably due to the star formation processes.
        The radius of the  T76 cluster is found to be 45$^{\prime}{^\prime}$ (1.24 pc) and 
	28 stars within this radius were marked as highly probable cluster members.
        We have found that the cluster is located at a distance of $5.7\pm1.0$ kpc and is
        having an age of $50\pm10$ Myr.
        The mass function slope ($\Gamma$) in the cluster region in the mass range $\sim$0.75$<$M/M$_\odot$$<$5.8 is estimated
        as $-1.3\pm0.2$, which is similar to the  value `-1.35' given by \citet{1955ApJ...121..161S}.
        The cluster is not showing any signatures of mass-segregation and is currently  undergoing dynamical relaxation.
\end{abstract}


}]


\doinum{12.3456/s78910-011-012-3}
\artcitid{\#\#\#\#}
\volnum{000}
\year{0000}
\pgrange{1--}
\setcounter{page}{1}
\lp{1}


	\section{Introduction}
	\label{sect:intro}

	As most of the stars form in a clustered environment in molecular clouds,
	the dynamics of stars in the clusters as well as the structure of clusters measured as a function of cluster age 
	hold important clues on the processes of star formation and stellar evolution \citep{2003ARA&A..41...57L}.  
	Many clusters show the distribution of massive stars towards their central region of clusters
	and whether this segregation of massive stars occurs due to an evolutionary effect 
	or is of primordial origin is not yet entirely clear. 
	
	\begin{figure*}
	\centering
	\hbox{
	\hspace{-0.5cm}
	\includegraphics[width=0.7\textwidth, angle=0]{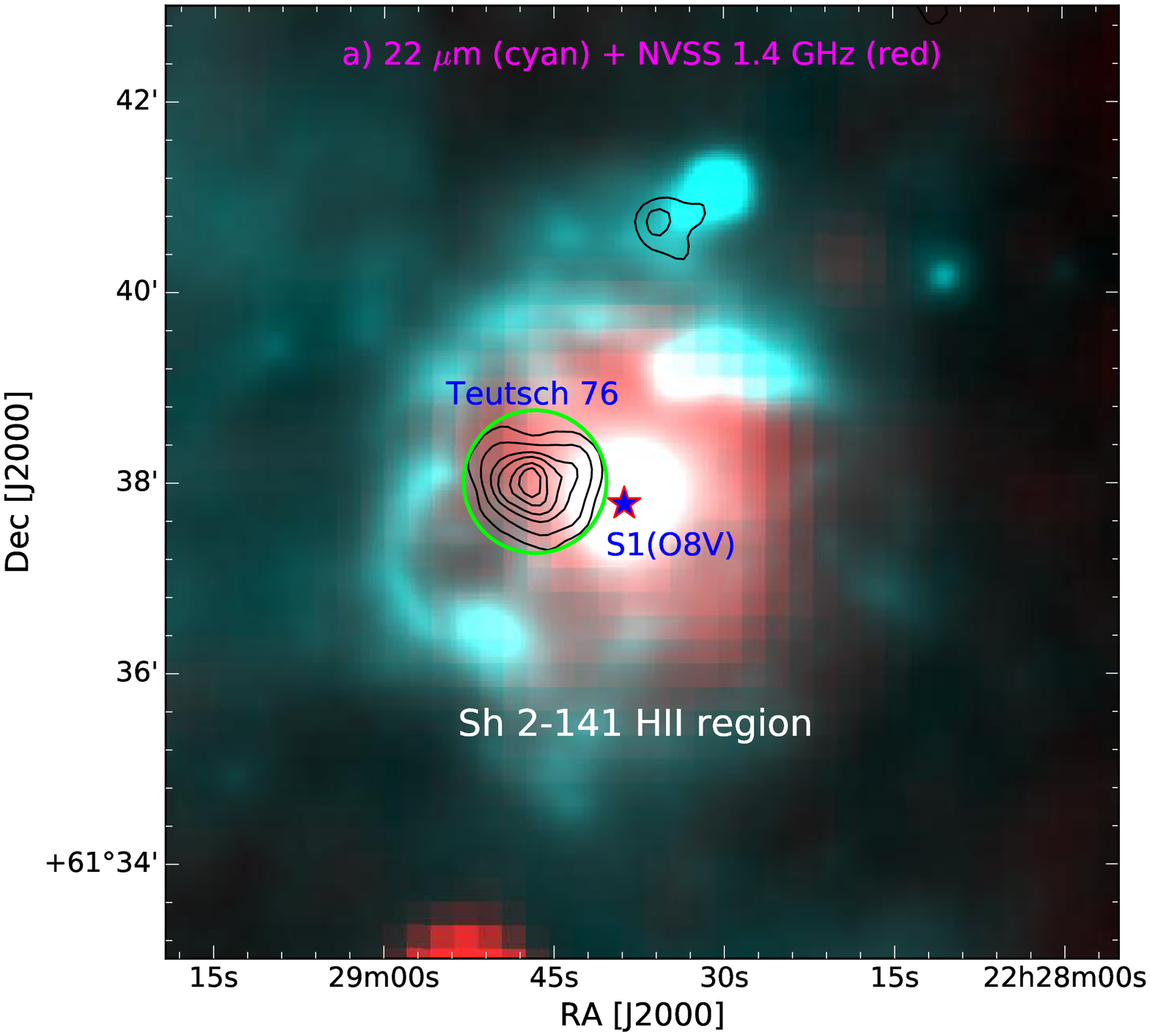}
	\hspace{-6.5cm}
	{\vbox 
	{
	\includegraphics[width=0.3\textwidth, angle=0]{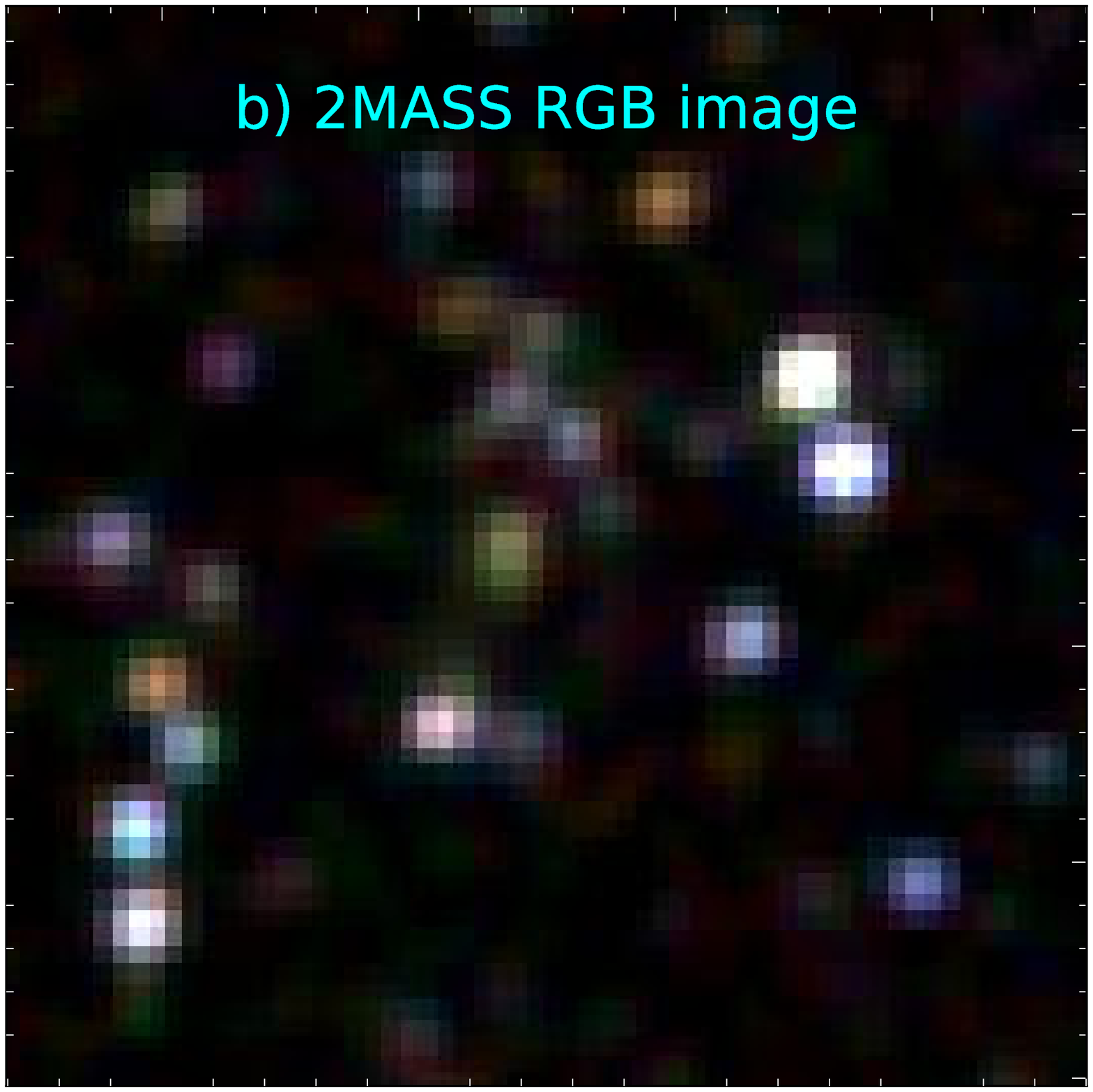}
	{\vbox{
	\vspace{0.6cm}
	\includegraphics[width=0.3\textwidth, angle=0]{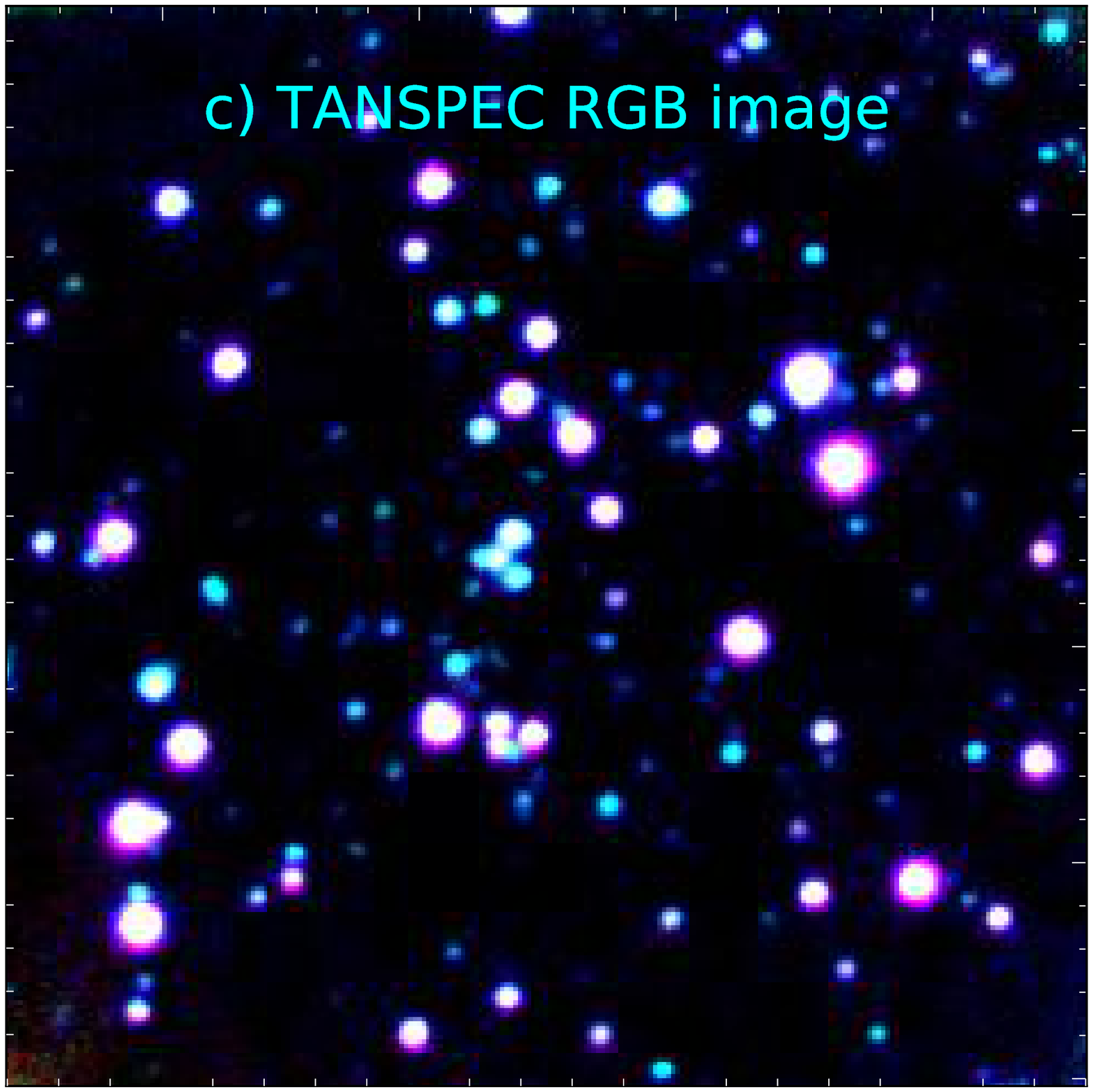}
	}}}}}
	\caption {
	(Left panel): Color-composite image obtained using the NVSS 1.4 GHz (red)
	and  WISE 22 $\mu$m  (cyan) images for an area of $\sim15\times 15$ arcmin$^2$ around the T76 cluster.
	The black contours are the isodensity contours generated using nearest 
	neighbor method from the 2MASS data (cf. Section 3.1).The green circle encloses the cluster T76 region. 
	(Right panels): Comparison of the color-composite images obtained by using the $J$ (blue), $H$ (green),
	and $K$ (red) images of the T76 cluster from the 2MASS (top panel) 
	and TANSPEC observations (bottom panel).
	}
	\label{2c}
	\end{figure*}

	Most of  the studies related to the stellar evolution and dynamics on star clusters during the past decade are not always based on deep photometric data 
	and lack the membership determination based on high-quality proper motion (PM) data.
	Teutsch 76 open cluster  ($\alpha_{J2000}$: $22^h28^m44^s.2$, $\delta_{J2000}$: $+61^\circ37^\prime52^{\prime \prime}$ \citep{2006A&A...447..921K} hereafter T76, cf.  Figure \ref{2c}),
	one of the poorly studied open clusters, is located in the Galactic plane towards 
	the 2$^{nd}$ Galactic quadrant ($l$ = 106$^\circ$.8171, $b$ = +03$^\circ$.3082).
	This cluster is located in the eastern part of the Sharpless region `Sh 2-141' inside a mid-infrared (MIR) bubble seen in WISE MIR band image (cf. Figure \ref{2c}).
	The Sh 2-141 H\,{\sc ii} region is reported to be ionised by an O8V star (named as `S1' hereafter) \citep{2007AA...470..161R}.
	We have performed a detailed analysis on this cluster to understand its dynamical evolution
	by using our deep near-infrared (NIR) observations taken from the TIFR-ARIES Near-infrared Spectrometer  \citep[TANSPEC;][]{2022PASP..134h5002S}
	recently installed on the 3.6m telescope at Devasthal, Nainital, India \citep{2018BSRSL..87...29K}, along with
	the recently available data from  the {\it Gaia} data release 3 \citep{2016A&A...595A...1G,2018A&A...616A...1G}
	and Pan-STARRS1 \citep{2016arXiv161205560C}.

	In this paper, Section \ref{sect:obs} describes the observations and data reduction.
	The structure of this cluster, membership probability of stars in the cluster region, 
	fundamental parameters (i.e., age and distance) of the cluster, and mass function (MF) analyses
	are presented in Section \ref{sect:result}.
	The dynamical structure of this cluster is discussed in Section \ref{sect:diss}, and
	we conclude our studies in Section \ref{sect:conclusion}.

	\begin{figure*}
	\centering
	\includegraphics[width=0.9\textwidth, angle=0]{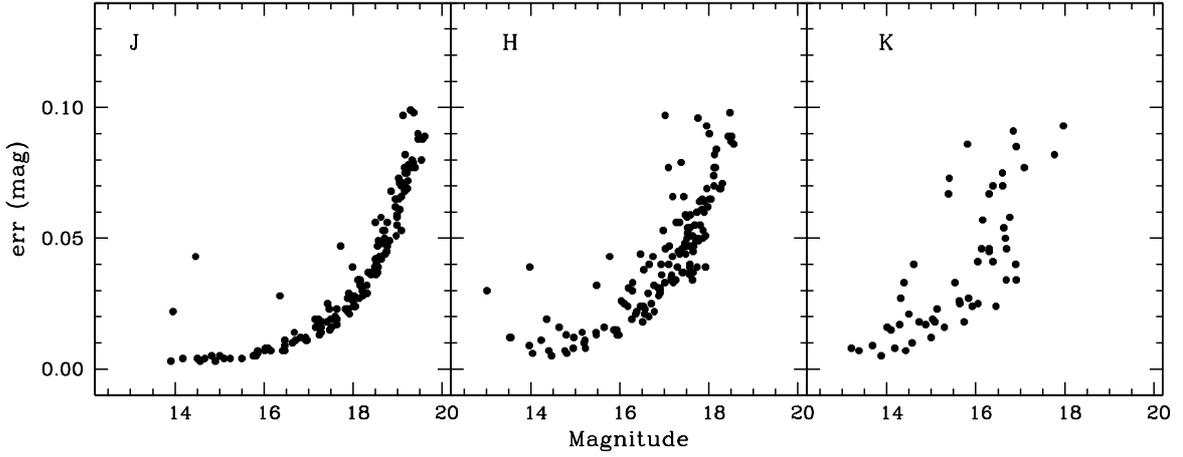}
	\caption {DAOPHOT errors as a function of $J,H,$ and $K$ magnitudes.}
	\label{fig:calibrate}
	\end{figure*}

	\section{Multi-wavelength  data sets}
	\label{sect:obs}

	\subsection{Deep NIR data}

	{\bf The central region of the T76 open cluster (refer Section 3.1)} was observed in the NIR  $J$ (1.20 $\mu$m), $H$ (1.65 $\mu$m) and
	$K$ (2.19 $\mu$m) bands (cf. Figure \ref{2c}) during the nights of 2020 November 19 and 27 using the
	TANSPEC instrument  mounted at the 
	Cassegrain main port of the 3.6m ARIES Devasthal Optical Telescope (DOT). 
	The weather conditions in these nights were good with relative humidity $<50$ percent
	and the full-width at half maxima of the stellar images was typically $\sim$0.7 arcsec in $J$ band.
	The field of view (FOV) of the TANSPEC is $\sim60\times60$ arcsec square with a
	plate scale of 0.244 arcsec.
	{\bf The observations were taken in 7 dither position with 135 frames, each having 20 secs of exposure.
	Thus, the total exposure time was 45 mins in each of the bands.}	
	Dark and sky flats were also taken during the observations.
	Sky frames in each filter were generated by  median combining the dithered frames.

 	The  basic data reduction including image cleaning, photometry and astrometry, is done using the  standard procedure 
	explained in \citet{2020MNRAS.498.2309S}. We transformed our instrumental $JHK$ magnitudes into a standard Vega system by using the following transformation equations.

	\begin{equation}
	(J-H)= (0.94\pm 0.11)\times(j-h) - (0.02\pm 0.08)
	\end{equation}
	\begin{equation}
	(J-K)= (0.78\pm 0.08)\times(j-k) - (0.51\pm 0.14)
	\end{equation}
	\begin{equation}
	(J-j)= (-0.03\pm 0.06)\times(J-H)- (2.54\pm 0.03)
	\end{equation}

	where the capital $JHK$ are the standard magnitudes of the
	stars taken from the 2MASS catalog and  the 
	small $jhk$ are the present instrumental magnitudes of the same stars normalized per sec exposure time.
	The DAOPHOT errors as a function of corresponding standard magnitudes are
	shown in Figure \ref{fig:calibrate}. 
	We have used only those stars for further analyses which are having
	signal-to-noise ratio greater than 10 (photometric errors $<$ 0.1 mag).
	In total, 143 stars were identified in the T76 cluster with detection limits of
	19.6 mag, 18.6 mag, 18.0 mag in $J$, $H$, $K$ bands, respectively.
	Figure \ref{2c} shows the comparison of a 2MASS image with the TANSPEC image. We can clearly see the resolved and faint stars
	in TANSPEC observations.
	Some of the brighter stars (3 in total) were saturated in our observations; we have taken their respective magnitudes from
	the 2MASS point source catalog.

	\subsection{Archival data}
	\label{sect:obs1}

	In order to study a wider area around T76, we have selected a FOV of $15\times15$ arcmin square
	as shown in Figure \ref{2c} and downloaded the available data from different survey, i.e.,
	{\it Gaia}  DR3\footnote{https://gea.esac.esa.int/archive/} \citep{2016A&A...595A...1G,2018A&A...616A...1G},
	The Panoramic Survey Telescope and Rapid  Response System (Pan-STARRS1 or PS1) data release 2 \footnote{http://catalogs.mast.stsci.edu/}  \citep{2016arXiv161205560C}, 
	and the 2MASS point source catalog\footnote{http://tdc-www.harvard.edu/catalogs/tmpsc.html} \citep{2003yCat.2246....0C}. 
	For our analyses, we have used only those sources which have photometric uncertainties less than 0.1 mag.

	\section{Results and Analysis}
	\label{sect:result}

	\subsection{Structure of  the T76 cluster}

	To study the structure of the T76 open cluster, we obtained stellar number density maps
	for the sample of stars taken from the 2MASS survey covering $15\times15$ arcmin 
	square FOV around this cluster region. 
	The stellar number density maps were generated using the nearest neighbor (NN) method as
	described by  \citet{2005ApJ...632..397G}. We took the radial
	distance necessary to encompass the sixth nearest stars and
	computed the local surface density in a grid size of 5 arcsec \citep[cf.][]{2009ApJS..184...18G}. 
	The stellar number density contours derived by this method are plotted in Figure \ref{2c} as black curves.
	The lowest contour is 1$\sigma$ above the mean of stellar density (13 stars/arcmin$^{2}$) 
	and the step size is equal to the 1$\sigma$ (3.5 stars/arcmin$^{2}$).
	As can be seen from the contours, the cluster is almost circular and is located within the Sh 2-141 H\,{\sc ii} region near a massive star
	S1 (O8V) \citep[][cf. Figure \ref{2c}]{2015AJ....150..147F,2007AA...470..161R}.
	The approximate boundary of the T76 cluster is shown with a green circle in the Figure \ref{2c}.
	The radius of the  T76 cluster is found to be 45$^{\prime}{^\prime}$ centered at
	 $\alpha_{2000}$: 22$^{h}$28$^{m}$46$^{s}$.68, $\delta_{J2000}$: +61$^\circ$38$^\prime$01$^\prime{^\prime}$.2  (cf. Figure \ref{2c}).

\begin{figure*}
\centering
\includegraphics[width=7.5cm,height=9cm]{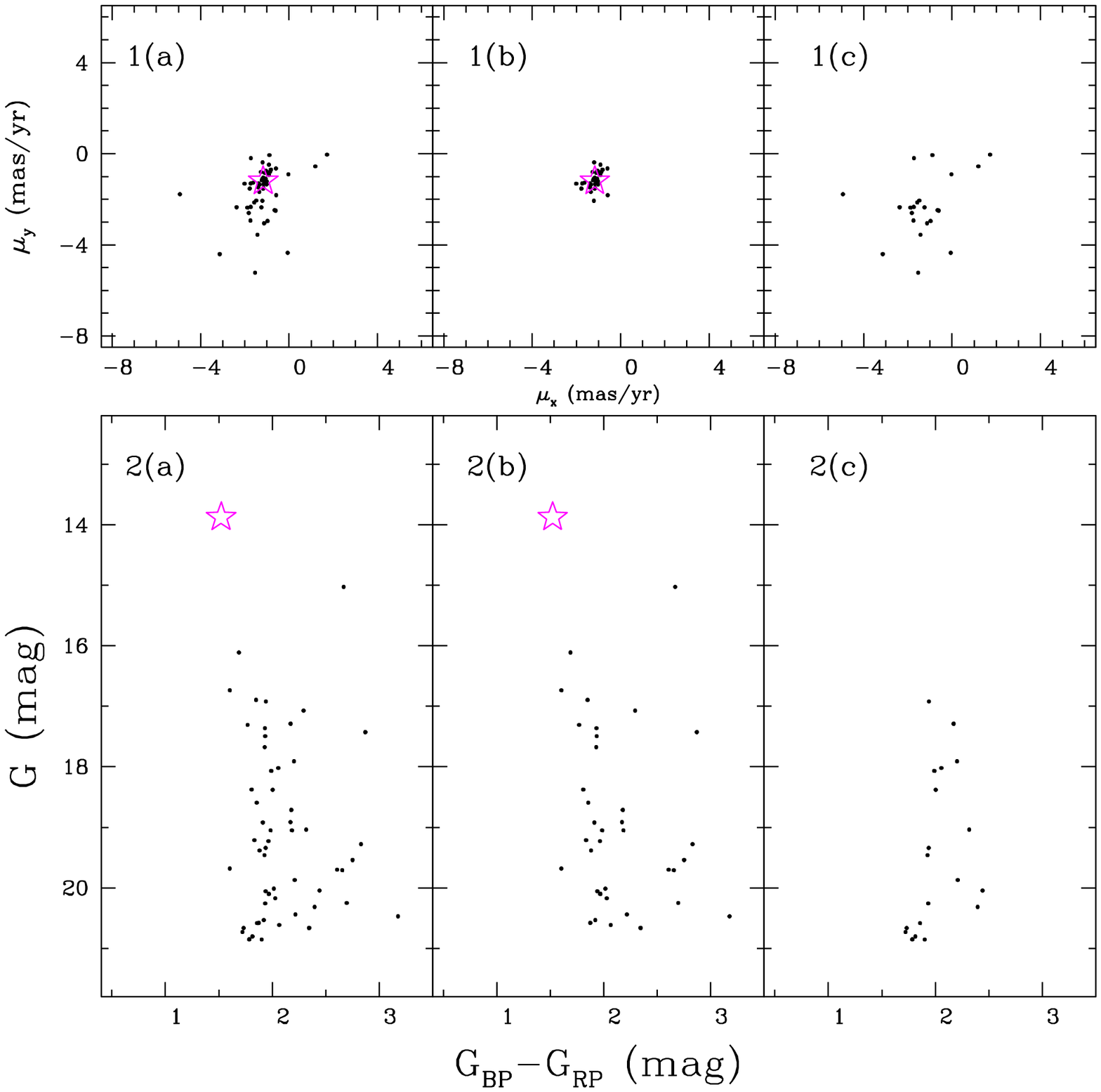}
\includegraphics[width=8cm,height=9cm]{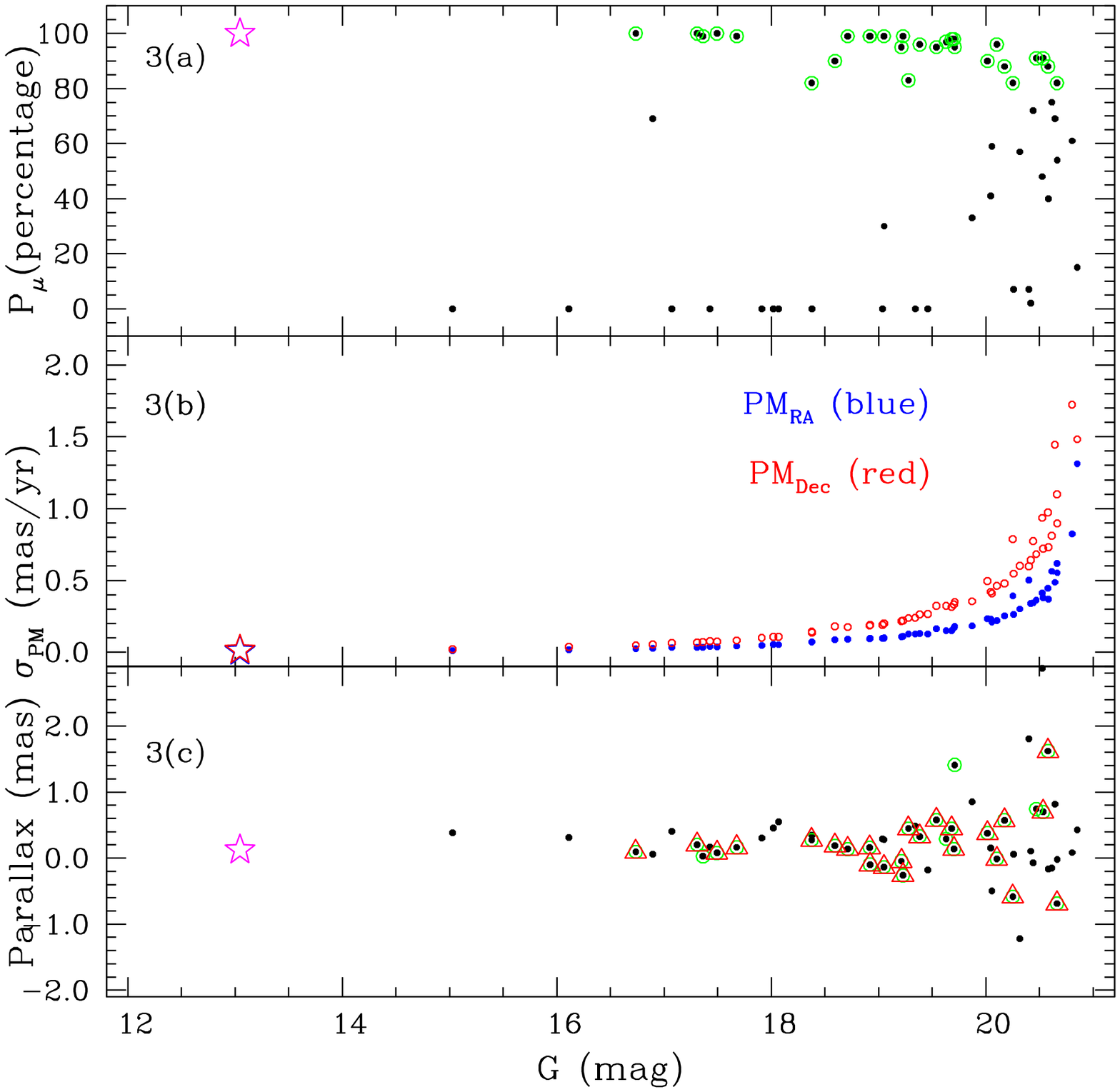}
\caption{\label{pm0} PM vector-point diagrams (VPDs; panel-1) and
$G$ vs. $(G_{BP}-G_{RP})$ CMDs (panel-2) for the stars located inside the T76 cluster region (cf. Section 3.1, radius $<$ 45$^{\prime}{^\prime}$).
The left sub-panels show all stars, while the middle and right sub-panels show the probable cluster members and field stars.
Panel-3: Membership probability P$_\mu$, PM errors $\sigma_{PM}$ and parallax of stars as
a function of $G$ magnitude for stars in the cluster region. The probable member stars (P$_\mu>$80 \%) are shown by green circles while the 24 members 
used for distance estimation of the T76 cluster are shown by red triangles (see text for details). Location of massive star S1 (O8V) is also shown in all panels by a star symbol.
}
\end{figure*}

	\subsection{Membership Probability}

Gaia DR3  has opened up the possibility of an entirely new perspective on the problem of
membership determination in cluster studies by
providing the new and precise parallax measurements upto very faint limits\footnote{https://gea.esac.esa.int/archive/}  \citep{2016A&A...595A...1G,2018A&A...616A...1G}.
$Gaia$  proper motion (PM) data located within the cluster region (cf. Section 3.1, radius $<$ 45$^{\prime}{^\prime}$)
and having PM error $\sigma_{PM}<$3 mas/yr
are  used to determine membership probability of stars located in this region.
Proper motions (PMs), $\mu_x$, i.e., $\mu_\alpha$cos($\delta$) and $\mu_y$ i.e., $\mu_\delta$, are plotted as vector-point diagrams (VPDs) in the panel-1 of
Figure \ref{pm0} (left panel). The panel-2 show the corresponding
$G_{(330-1050 nm)}$ versus $G{_{BP (330-680 nm)}} - G_{{RP(630-1050 nm)}}$
Gaia color-magnitude diagrams (CMDs).
The left sub-panels show all stars, while the middle and right sub-panels show
the probable cluster members and field stars. A circular area of
a radius of 1 mas yr$^{-1}$ {\bf (keeping in mind the errors  and the expected dispersion in the PM of cluster  stars)} 
around the cluster centroid in the VPD of PMs has been selected visually to define
our membership criterion. The chosen radius is a compromise between
losing cluster members with poor PMs and including field
stars sharing mean PM.  
The CMD of the most probable
cluster members are shown in the lower-middle sub-panel.
The lower-right sub-panel represents the CMD for field stars. Few cluster members are visible in
this CMD because of their poorly determined PMs.
The tight clump centering at $\mu_{xc}$ = -2.4 mas yr$^{-1}$, $\mu_{yc}$ = -1.1 mas yr$^{-1}$ and radius = 1 mas yr$^{-1}$
in the top-left sub-panel represents the cluster stars, and a broad distribution is seen for the probable
field stars.
Assuming a distance of $\sim$ 5 kpc (cf. Section 3.3) and a
radial velocity dispersion of 1 kms$^{-1}$ for open clusters \citep{1989AJ.....98..227G}, the expected dispersion ($\sigma_c$) in PMs of the cluster would be $\sim$0.04 mas yr$^{-1}$.
For remaining stars (probable field stars), we have calculated:
 $\mu_{xf}$ = -2.65 mas yr$^{-1}$, $\mu_{yf}$ = -2.14 mas yr$^{-1}$,
$\sigma_{xf}$ = 2.53 mas yr$^{-1}$ and $\sigma_{yf}$ = 1.29 mas yr$^{-1}$.
These values are further used to construct the frequency distributions of cluster stars ($\phi_c^{\nu}$) and field stars ($\phi_f^{\nu}$) by using the equations given in \citet{2013MNRAS.430.3350Y} and then the value of
membership probability (ratio of distribution of cluster stars with all the stars) of all the stars within the T76 cluster (Section 3.1, radius$<$ 45$^\prime{^\prime}$), 
is given by using the following equation:

\begin{equation}
P_\mu(i) = {{n_c\times\phi^\nu_c(i)}\over{n_c\times\phi^\nu_c(i)+n_f\times\phi^\nu_f(i)}}
\end{equation}

where $n_c$ (=0.55) and $n_f$(=0.45) are the normalized number of stars for the cluster
and field ($n_c$+$n_f$ = 1), respectively. The membership probability estimated as above,
errors in the PM, and parallax values are plotted as a function of
$G$ magnitude in panel-3 of Figure \ref{pm0}.
As can be seen in this plot, a high membership probability (P$_\mu >$ 80 \%)
extends down to $G\sim$20 mag.
At brighter magnitudes, there is a  clear separation between cluster
members and field stars supporting the effectiveness of this technique. Errors in PM become very large at faint limits and
the maximum probability gradually decreases at those levels.
Except few outliers, most of the  stars with high membership probability (P$_\mu >$ 80 \%) are
following a tight distribution.
Finally, from the above analysis, 28 stars were considered as  members of the T76 cluster
based on their high membership probability P$_\mu$ ($>$80 \%).

	\begin{figure}
	\centering
	\includegraphics[width=0.23\textwidth, angle=0]{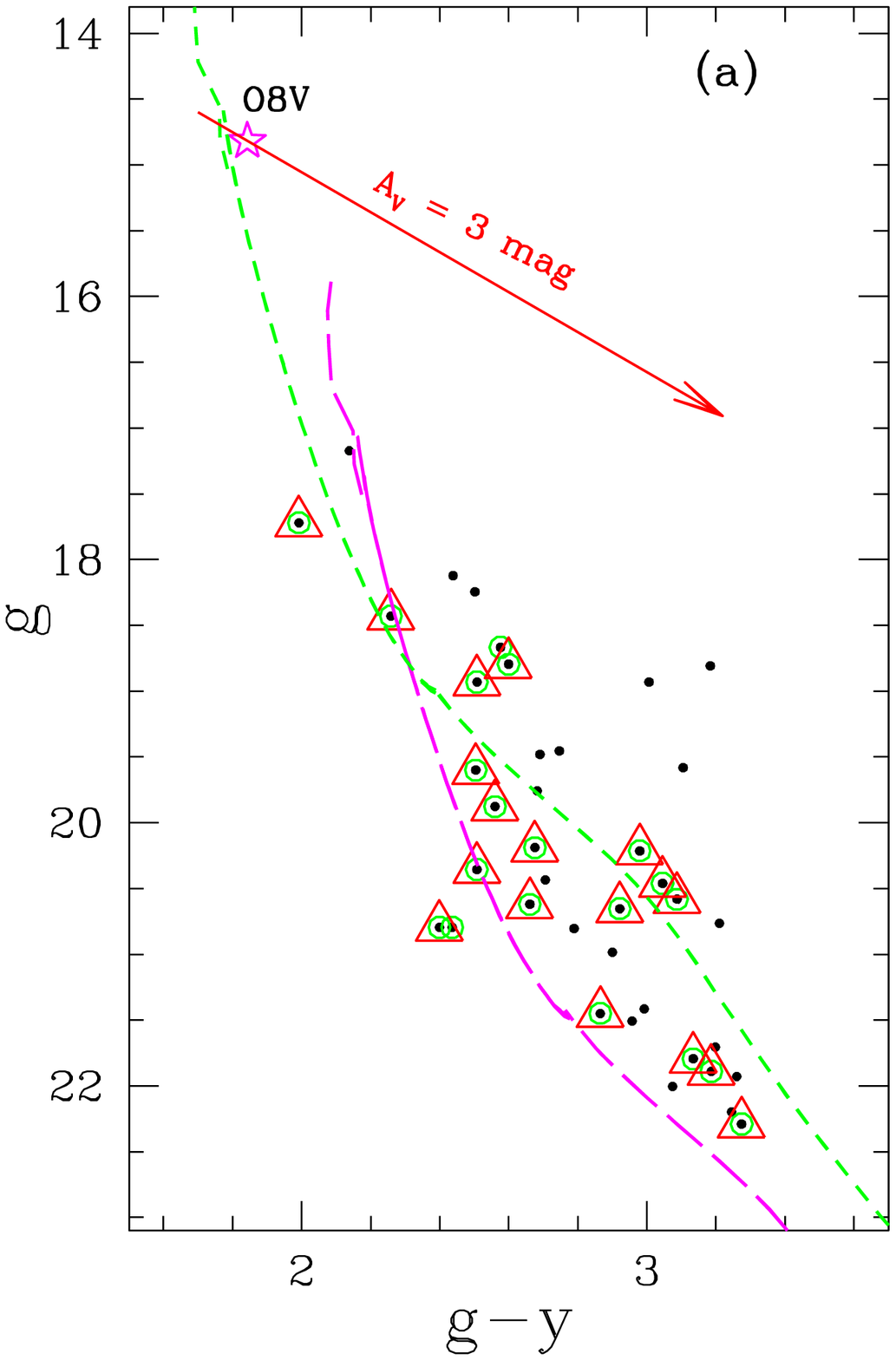}
	\includegraphics[width=0.23\textwidth, angle=0]{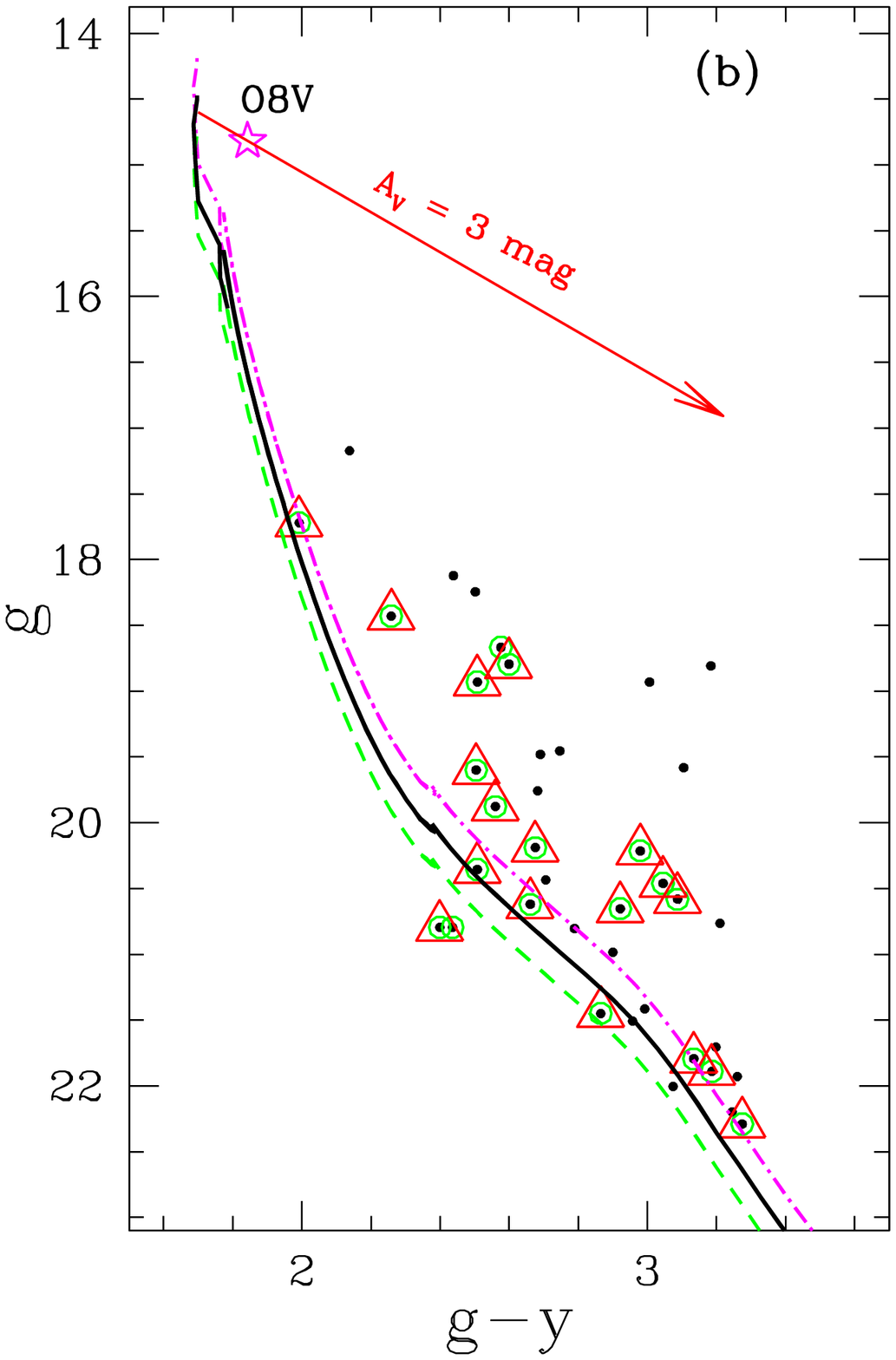}
	\caption {(a) PS1 $g$ vs. $(g - y)$ CMD for the stars  in the cluster region (black dots).
	The green circles are the member stars of the cluster.
	The curves denote a ZAMS derived from \citet{2019MNRAS.485.5666P}
		corrected for extinction and distance  values reported earlier, i.e.,  \citep[$A_V$=3.95 mag, distance = 3.5 kpc][green dashed curve]{2016AA...585A.101K} 
		and  \citep[$A_V$=4.46 mag, distance = 8.34 kpc][ magenta dashed curve]{2007AA...470..161R}. 
	 (b) Same as panel (a) but for CMD fitted with $A_V$=3.95 mag \citep{2019ApJ...887...93G,2016AA...585A.101K} and distance =  5 kpc (magenta curve, mean distance of cluster members), 
	6.4 kpc (green curve, distance of massive star `S1') and 5.7 kpc (black curve, CMD fitted value).   
	}
	\label{fgy1}
	\end{figure}
	
	\begin{table*}
	\centering
	\small
	
		\caption{\label{param} Parameters for T76/Sh 2-141.}
	\begin{tabular}{lcccccc}
	\hline
		Reference			& V$_{LSR}$& Kinematical   &Spectrophotometric & A$_V$& Ionizing   \\
						&(km/s)	   &Distance (kpc) &Distance (kpc)& (mag) &Star/Age \\
	\hline
		 This work			&$-$  &	$-$	  &$5.7\pm1.0$	&3.95	&50 Myr	\\
		\citet{2019ApJ...887...93G}	&$-$  &	$-$	  &$-$		&3.94	&$-$	\\
		\citet{2018ApJS..238...28K}	&$-$	   &	$-$	  &	$-$	&3.75	&$-$	\\
		\citet{2016AA...585A.101K}	&$-$	   &$-$		  &3.5		&3.95	&		 8 Myr\\
		\citet{2015ApJS..221...26A}	& -62.9 (X-band, 9GHz, 3cm)   & $7\pm1.2$ 	  &$-$		&$-$&$-$	\\
		\citet{2015AJ....150..147F}	& -65.8($^{12}$CO) $-$ -62.9(HI)&$-$&$9.92\pm1.98$&3.90& O8V \\
		\citet{2007AA...470..161R}	& -64.4	   &$-$		  &$8.34\pm0.6$ &4.46   & O8V		\\
		\citet{2000AJ....120.3218P}	& -65(CO$^1$) $-$ -63.8(H$\alpha^2$)  & 7 	 	  &$-$		&$-$	& O8		\\
	\hline
	\end{tabular}
	\tablenotes{$^1$: \citep{1984ApJ...279..125F}; $2$: \citep{1990AJ.....99..622F}}
	\end{table*}

	\begin{figure*}
	  \centering
	 \includegraphics[width=0.25\textwidth, angle=0]{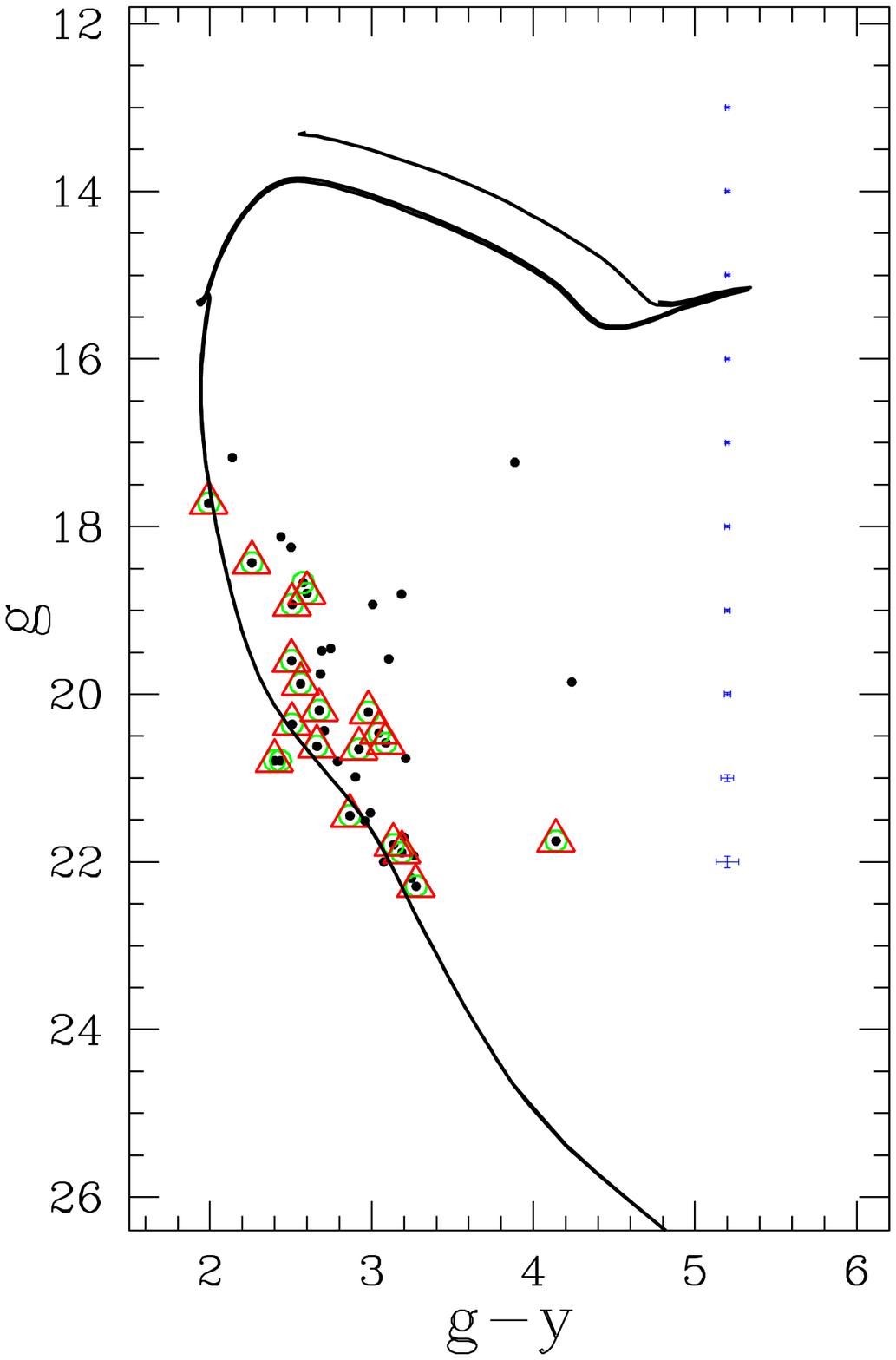}
	 \includegraphics[width=0.25\textwidth, angle=0]{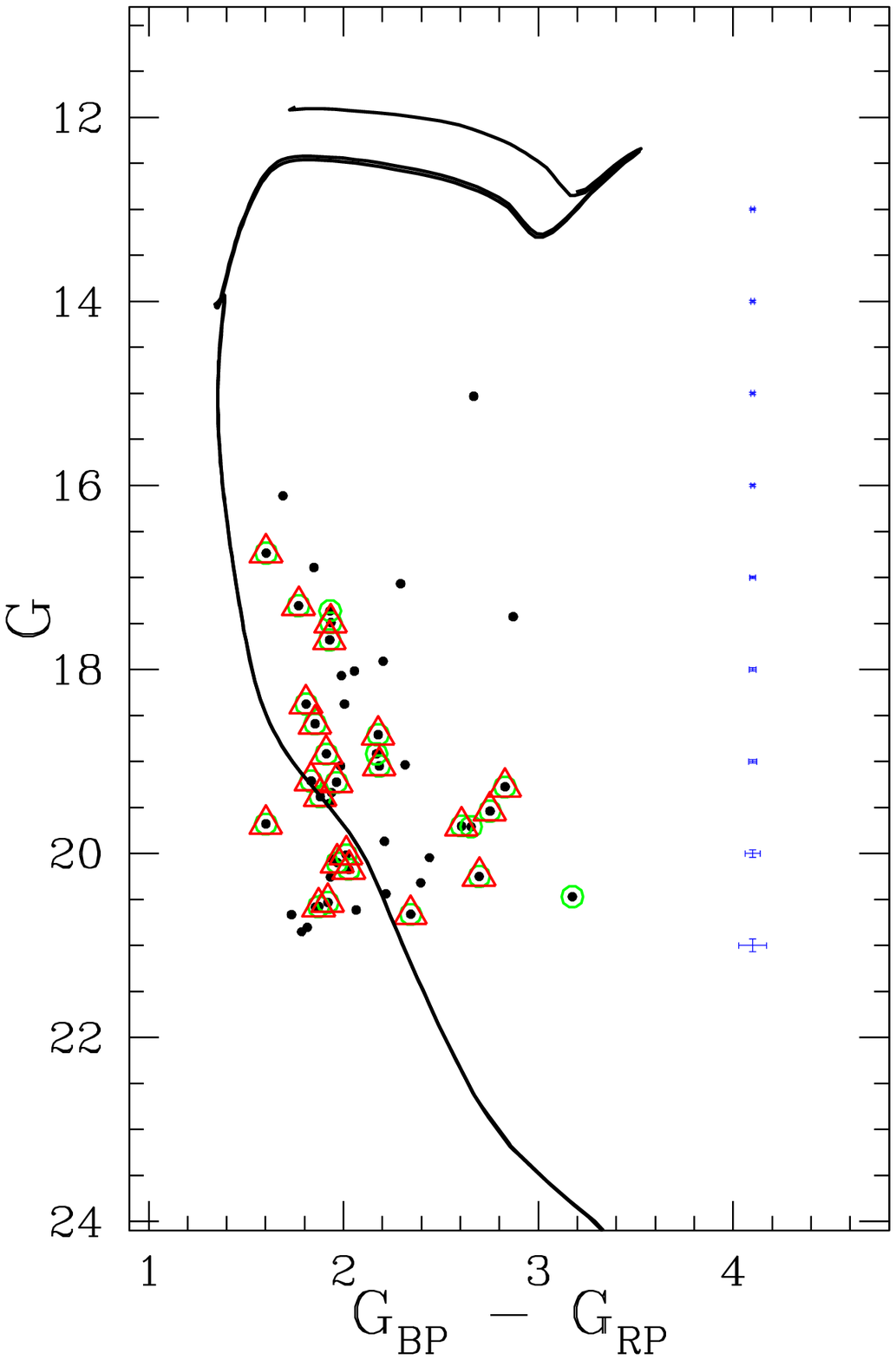}
	 \includegraphics[width=0.25\textwidth, angle=0]{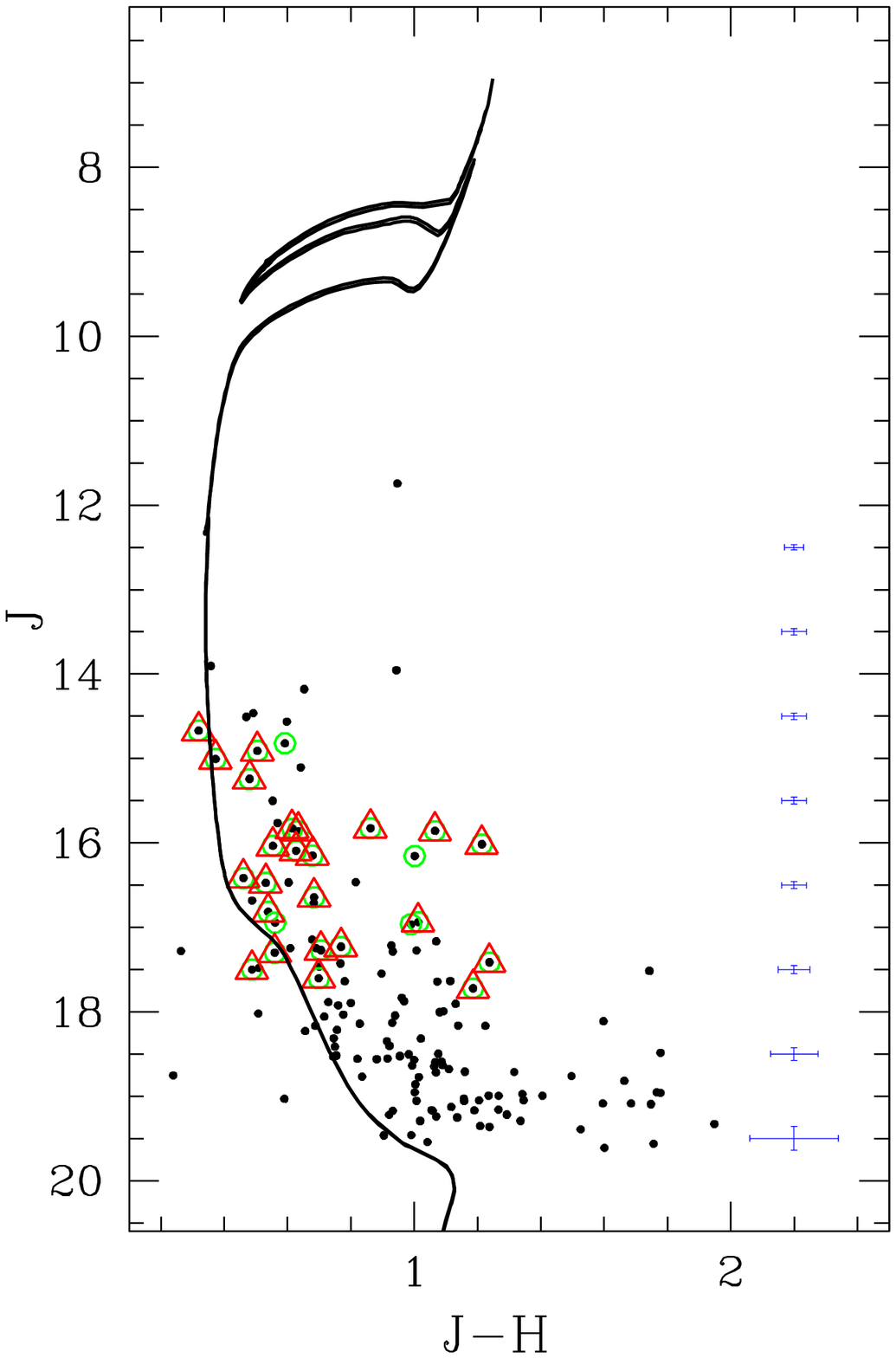}
		\caption {The CMDs generated from PS1 (left panel), $Gaia$ (middle panel), 
		and TANSPEC data (right panel) for the stars  in the cluster region.
		Symbols are the same as in Figure \ref{fgy1}.
		The black curve denotes an isochrone of age = 50 Myr  derived from \citet{2019MNRAS.485.5666P}.
		The isochrone is corrected for distance (5.7 kpc) and extinction ($A_V$=3.95 mag). 
		Photometric error bars are also shown in the CMDs.}
	  \label{fgy2}
	\end{figure*}

	\subsection{Distance and age of the cluster}

	We have calculated the mean of the reported photo-geometric distances of 24 members of the T76 cluster \citep{2021AJ....161..147B} 
	(leaving a couple of outliers, as shown in Figure \ref{pm0} with red triangles) as $\sim$5$\pm$1 kpc.
	The previous spectro-photometric measurements 
	place this cluster at  different distances ($\sim$3.5 - 10 kpc, cf. Table \ref{param}).
	The ionizing source of the Sh -141 H\,{\sc ii} region, i.e.,  `S1', an O8V star, is also located at a farther
	distance of 6.4$^{+0.8}_{-0.4}$ kpc  \citep{2021AJ....161..147B}. 
	The reported kinematical distance of the molecular cloud containing the T76 cluster and the Sh 2-141 
	H\,{\sc ii} region is $7\pm1.2$ \citep{2015ApJS..221...26A,2000AJ....120.3218P}. 
	The distances to clusters with mean parallaxes smaller than $\sim$0.2 mas (distance $\geq$5 kpc) are better constrained by
        classical isochrone fitting methods \citep[e.g.,][]{2022ApJ...926...25P, 2020MNRAS.492.2446P,2020ApJ...891...81P, 2020MNRAS.498.2309S, 2017MNRAS.467.2943S,2006AJ....132.1669S, 2018A&A...618A..93C, 1994ApJS...90...31P}.
	Therefore, to further check the validity of PM distance estimation of T76 ($\sim$5$\pm$1 kpc), we have used the 
	PS1 $g$ vs. $(g - y)$ CMD for the stars in the cluster region along with  the
        member stars as shown in Figure \ref{fgy1}a.
	The index $(g-y)$ has been used here because of having a very large color range.
        We have also shown the ZAMS derived from \citet{2019MNRAS.485.5666P}
        corrected for extinction and distance  values reported earlier, i.e.,  \citep[$A_V$=3.95 mag, distance = 3.5 kpc;][]{2016AA...585A.101K}
        and  \citep[$A_V$=4.46 mag, distance = 8.34 kpc;][]{2007AA...470..161R}.
	Clearly, both sets of these parameters do not fit to cluster stars distribution in the CMD.
        In Figure \ref{fgy1}b, we show similar CMD but fitted with $A_V$=3.95 mag \citep{2019ApJ...887...93G,2016AA...585A.101K} 
	and distance =  5 kpc (mean distance of cluster members), 6.4 kpc (distance of a massive star `S1') and 5.7 kpc (black curve, CMD best fitted value).
	The distance of 5.7 kpc is estimated from the visually fit of ZAMS to the
	lower envelope of the distribution of member stars where the bend occurs in the MS
	\citep[see for details,][]{1974ASSL...41.....G,1994ApJS...90...31P}.
	Clearly, out of these three ZAMSs corrected for  different distance estimates, 
	the member stars are best represented by a ZAMS corrected for a distance of 5.7 kpc.
	The massive star `S1' also represented best by this distance estimate as it can be traced back to ZAMS along with reddening vector to the intrinsic color of 
	O8V spectral type star (cf. Figure \ref{fgy1}b).
	Here, it is worthwhile to note that the visual fitting in this case is prone to large error
	as the cluster can have differential reddening due to nebulosity around it.
	Therefore, we have estimated the error in the distance as 1.0 kpc using the procedure outlined in \citet{1994ApJS...90...31P}.

	Since the T76 cluster seems to be associated with a H\,{\sc ii} region, there might be a probability of finding young stars in it with excess IR emission.
	We have tried to find them using the conventional NIR color based selection criteria \citep{2007MNRAS.380.1141S} but have found none.
	Therefore, to derive the age  of the T76 cluster, we have used the deep multi-wavelength data from 
	present observations, $Gaia$, and PS1, to generate
	CMDs in different color spaces are shown in  Figure \ref{fgy2}.
	The CMDs  display a  well defined MS and a MS turn-off point.
	We can visually fit an isochrone of age $\sim$50 Myr (solid black curve) 
	taken from  \citet{2019MNRAS.485.5666P} to the distribution of stars in the post-MS phase in all the CMDs. 
	We are expecting 20\% error in this age estimation \citep[see e.g.,][]{1994ApJS...90...31P}.	
	
	From the above analysis, it seems that the cluster T76 is located at a farther distance of $5.7\pm1.0$ kpc and is having an age of $\sim50\pm10$ Myr.

	\subsection{Mass function}

\begin{figure}
	  \centering
	  \includegraphics[width=0.45\textwidth, angle=0]{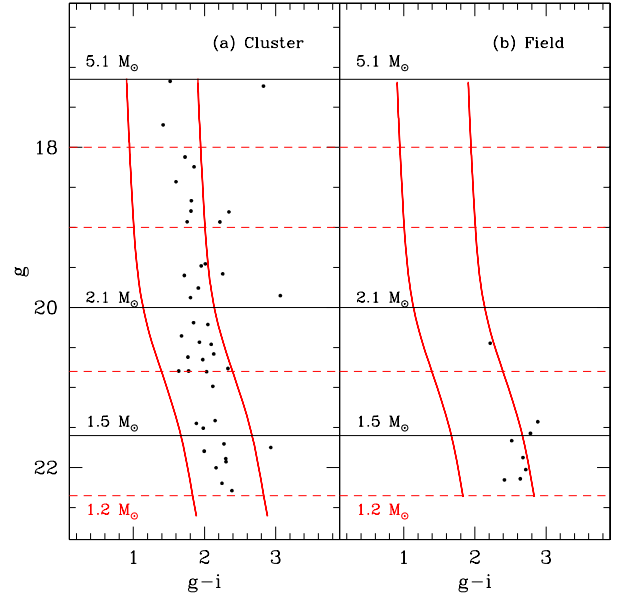}
		\caption {PS1 $g$ vs. $(g - i)$ CMDs for the stars  in the (a) cluster 
		and (b) field regions.  
		The curves denote a MS envelope created by 
		the MS isochrone of 50 Myr derived from \citet{2019MNRAS.485.5666P} corrected for the
		distance (5.7 kpc) and  {\bf extinction ($A_V$=3.95 mag) (see text for details)}.
		Upper and lower horizontal lines represent the MS turn-off point and 50 percent completeness limit, respectively.
	}
	  \label{fband}
	\end{figure}

	Open clusters possess many favorable characteristics for
	MF studies, e.g., clusters contain almost coeval set of
	stars at the same distance with the same metallicity; hence,
	difficulties such as complex corrections for stellar birth rates,
	life times, etc, associated with determining the 
	MF from field stars are automatically removed. 
	The MF is often expressed by a power law,
	$N (\log m) \propto m^{\Gamma}$ and  the slope of the MF is given as:

	\begin{equation}
	\Gamma = d \log N (\log m)/d \log m 
	\end{equation}

	where $N (\log m)$ is the number of stars per unit logarithmic mass interval.
	The MS luminosity function (LF) obtained with the help of $g$ versus $(g-i)$ CMD generated from the 
	deep PS1 photometric data  (cf. Figure \ref{fband}) and corrected for the data incompleteness, has been converted into an MF 
	using the isochrone of \citet{2019MNRAS.485.5666P} of age $\sim$50 Myr, 
	corrected for the distance and extinction \cite[see also][and references therein]{2020MNRAS.498.2309S}.

        \begin{figure}
          \centering
          \includegraphics[width=0.45\textwidth, angle=0]{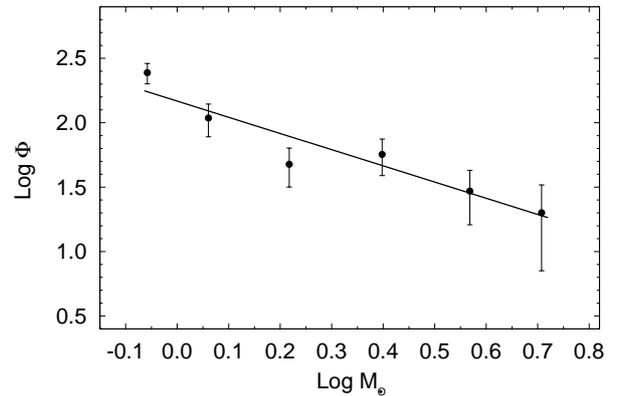}
                \caption {A plot of the MF for the cluster region of T76 using PS1 data.
        Log $\phi$ represents log($N$/dlog $m$). Open circle represents data below 50 percent completeness limit
        and is not used in the MF analysis.
                The error bars represent $\pm\sqrt N$ errors.
        The solid line shows the least squares fit to the MF distribution (black filled circles).
        }
          \label{mf1}
        \end{figure}

	\begin{figure}
	\centering
	\vbox{\includegraphics[width=0.48\textwidth, angle=0]{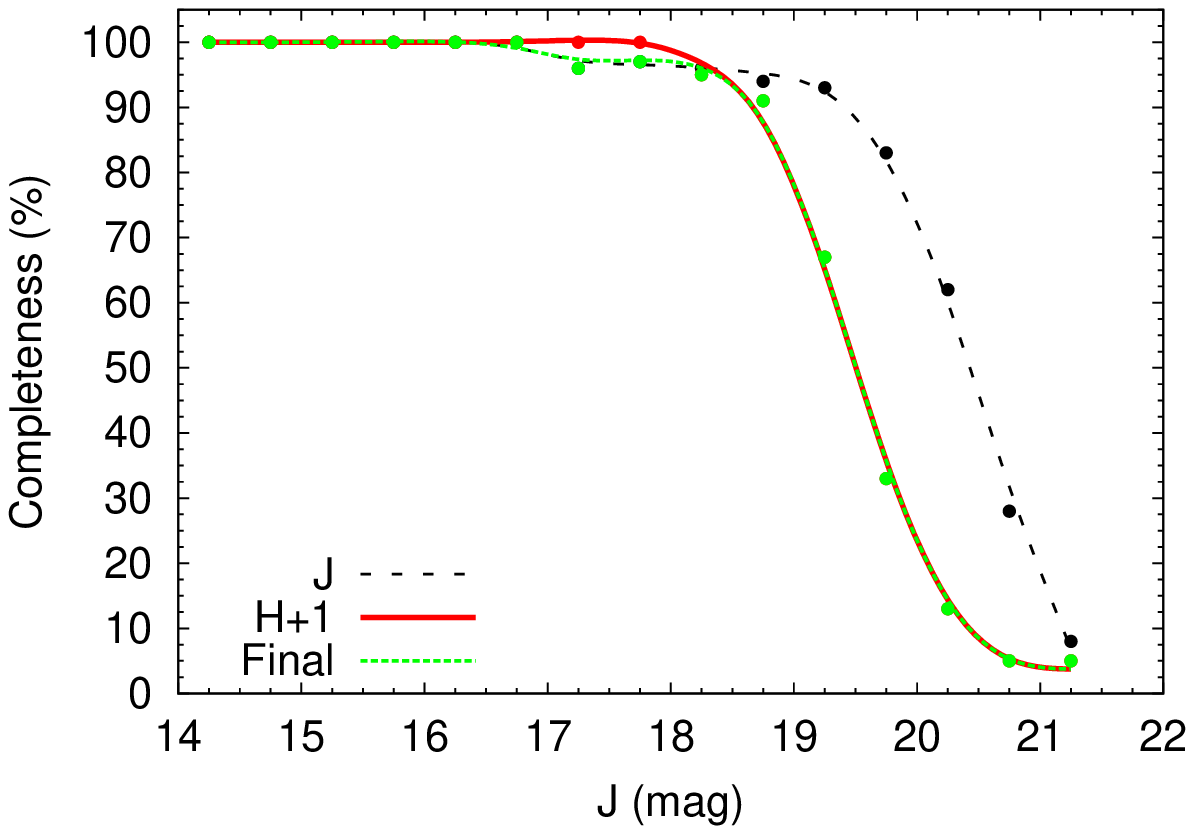}}
	\vbox{\includegraphics[width=0.48\textwidth, angle=0]{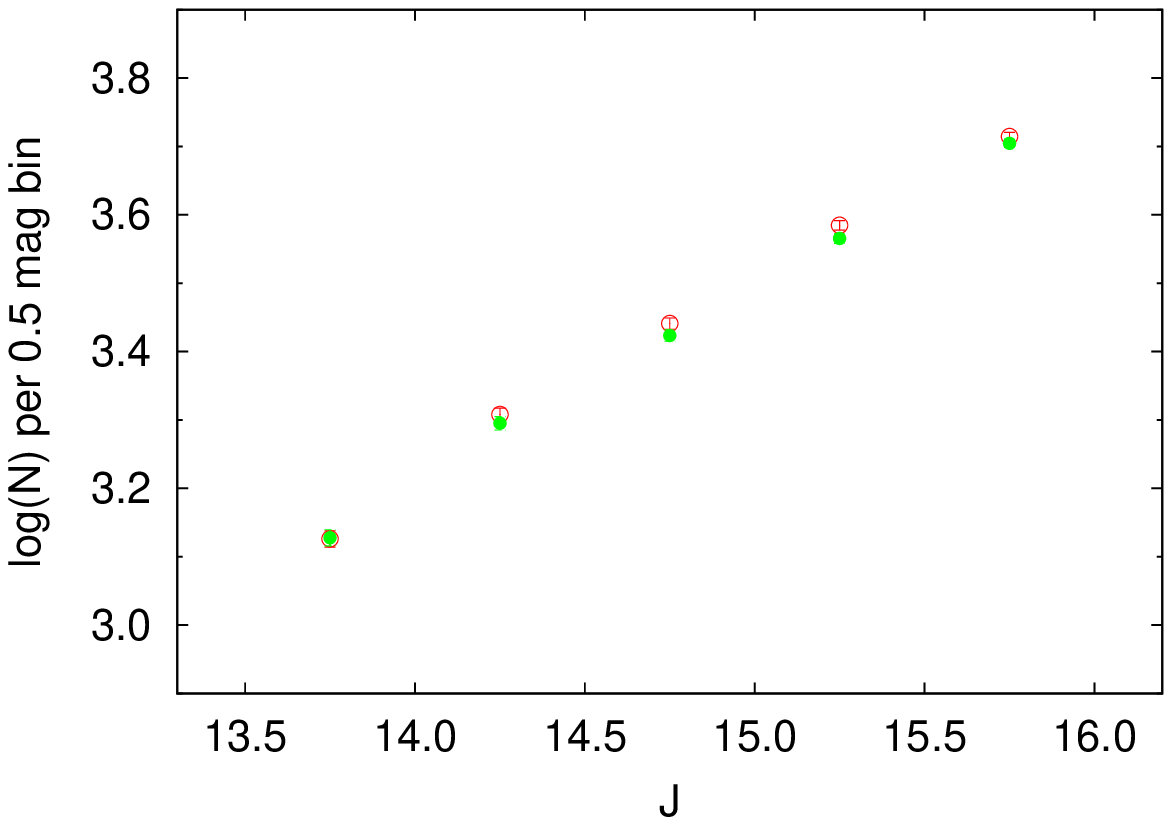}}
	\vbox{\includegraphics[width=0.48\textwidth, angle=0]{mf.eps}}
	\caption {(Top panel): The completeness factor as a function of $J$ magnitude
	derived from the artificial star experiments ({\it ADDSTAR}, see   \citet{2020MNRAS.498.2309S} for details)
	on the TANSPEC $J$ and $H$ band images.
	The $H$-band completeness factor is off-setted by the mean color of the MS stars (i.e., 1.0 mag).
	The continuous curves are the smoothened bezier curves for the data points for completeness.
	(Middle panel): A comparison of field stars distribution generated by using a nearby reference 
	field (green filled circles)  and by the model/simulations generated by the Besan\c con model (red open circles).
	(Bottom panel): A plot of the MF for the T76 cluster using TANSPEC data.
	Log $\phi$ represents log($N$/dlog $m$). The error bars represent $\pm\sqrt N$ errors. 
	The solid line shows the least squares fit to the MF distribution (black filled circles).
	}
	\label{mf2}
	\end{figure}

	In the Figure \ref{fband}, 
	we show the CMD for the cluster region as well as for the reference region ($\alpha_{J2000}$: $22^h28^m08^s.4$, $\delta_{J2000}$: $+61^\circ34^\prime40^{\prime \prime}.7$)
	having the same area. The contamination due to field stars is greatly reduced by selecting
	a sample of stars which are located near the well-defined MS \citep[cf.][]{2008AJ....135.1934S}. 
	Therefore, we generated an envelope of $\pm 0.5$ mag around the CMD keeping in mind the distribution of member stars and
	is shown in the left panel of Figure \ref{fband}. 
	As the MS is extended from $\sim$17.15 mag ($\sim$5.1 M$_{\odot}$) to $\sim$22.35 mag ($\sim$1.2 M$_{\odot}$), 
	the number of probable cluster members were obtained by subtracting 
	the contribution of field stars  (corrected for data incompleteness), in different magnitude bins having size of 1.0 mag
	from the contaminated sample of MS stars (also corrected for data incompleteness).
	We have use the estimation of the completeness factor (CF) for the PS1 data 
	as has been estimated in our previous paper, i.e., \citet{2020MNRAS.498.2309S}.
	The  photometric data is 90 percent complete upto 21.6 mag  
	in the $g$-band which corresponds to a star of mass 1.5 M$_\odot$ at the distance of the T76 cluster.
	We have also shown the MS turn-off point and 50 percent completeness limit in Figure \ref{fband}.
	The resultant MF distribution for the cluster 	region is shown in Figure \ref{mf1}.
	The slope of the MF ($\Gamma$) in the mass range $\sim$1.5$<$M/M$_\odot$$<$5.1
	comes out to be $-1.6\pm0.3$ for the stars in the  T76 cluster region.

	We have also used the present deep NIR photometry taken from the TANSPEC to derive the MF slope of the T76 cluster. 
	The CF is determined for $J$ versus $J-H$ CMD using the same procedure as discussed in  \citet{2020MNRAS.498.2309S}
	and is shown in the top panel of Figure \ref{mf2}.
	To decontaminate the field star population, we have used the CMD of a nearby reference field taken from 
	the 2MASS survey (for stars having $J<16$ mag)
	and the Besan\c con Galactic model of stellar population synthesis \citep{2003A&A...409..523R,2004ApJ...616.1042O} 
	(for stars having $J>16$ mag).
	To check the accuracy of statistics of number of stars generated by the Besan\c con model, 
	we have compared the LF generated from the model with that from the 2MASS survey ($J<16$ mag) 
	in the middle panel of Figure \ref{mf2}. 
	The LFs from both methods are matching well and the resultant MF distribution is
	shown in the bottom panel of Figure \ref{mf2}. 
	The value of the MF slope  ($\Gamma$) for the T76 cluster
	estimated by using the deep TANSPEC data  comes out to be $-1.3\pm0.2$ in the mass range $\sim$0.75$<$M/M$_\odot$$<$5.8.

	\section{Discussion}
	\label{sect:diss}

	Using our deep NIR data, the MF slope upto 0.75 M$_\odot$ is found to be $\Gamma=-1.3\pm0.2$ for the cluster region, 
	which is very similar (i.e., $\Gamma$=-1.35) to that reported by  \citet{1955ApJ...121..161S}.
	This indicates that the distribution of stars in this cluster is similar to the distribution found in our solar neighborhood 
	and the low mass stars are still intact with the cluster and there is no effect of dynamical evolution on them as of now.
	To further check this, we have used \citet{2009MNRAS.395.1449A} method for calculating mass segregation ratio (MSR)
	as a measure to identify and quantify mass segregation in the cluster.
	This method works by constructing the minimal sampling tree (MST) for massive stars and for the equal number of randomly selected stars from the cluster sample
	and estimating the ratio of their mean edge length, $\Gamma_{MSR}$  \citep[see for details,][]{2020MNRAS.498.2309S, 2018MNRAS.473..849D,2011A&A...532A.119O}.
	We have used the magnitudes of member stars (cf. Section 3.4) as a proxy for the mass. 
	This avoids uncertainties when we convert the observed luminosities into masses \citep{2018MNRAS.473..849D}.
	For the T76 cluster, we have estimated the value of $\Gamma_{MSR}$ as  $0.9\pm1.7$, 
	which dissuades the effect of mass-segregation in this cluster \citep[see also,][]{2020MNRAS.498.2309S}.
	A value of $\Gamma_{MSR}\approx 1$ implies that both
	samples of stars (i.e. the most massive and the randomly selected)
	are distributed in a similar manner, whereas  $\Gamma_{MSR} > 1$
	indicates mass segregation and $\Gamma_{MSR}\ll 1$ points to inverse mass segregation, 
	i.e. the massive stars are more spread outwards than the rest.

	To confirm the dynamical state of this cluster, we have estimated the dynamical
	relaxation time, $T_E$, the time in which the individual stars
	exchange sufficient energy so that their velocity distribution
	approaches that of a Maxwellian equilibrium, using the method given by \citet{1987gady.book.....B}.
	By counting the number of member stars (71 stars, cf. Section 3.4), 
	the value of $T_E$ comes out to be $\sim$12 Myr for the T76 cluster  \citep[see also,][]{2020MNRAS.498.2309S}.
	If we assume loss of 50\% of stars due to incompleteness of our data,
	the  dynamical relaxation time will be $T_E \sim$20 Myr, which is only 2.5 times less than that of the estimated age of the T76 cluster (50 Myr).
	This indicates that the T76 cluster is still under the process of dynamical relaxation.
	Usually the low-mass member stars become the most vulnerable
	to be ejected out of the system due to the dynamical relaxation, i.e, stellar evaporation happens with an e-folding time scale of
        $\tau_{evap}$ $\sim100\times T_{E}$ \citep{1982phyn.book.....S,1984ApJ...284..643M, 1987gady.book.....B}.
	As $\tau_{evap}$ comes out to be $\sim$1 Gyr for T76 having age $\sim$50 Myr, we can safely assume that the
	low mass stars have not ejected out of the cluster due to dynamical relaxation and the MF slope which we have estimated is very much representative of
	the primitive IMF of the clusters. 
	The typical survival time scale of open clusters in the Galactic
	disk is about 200 Myr \citep{2006A&A...446..121B,2013ApJ...762....3Y}.
	Open clusters much older than the survival timescale usually have
	distorted shape and loosened structure which leads to their
	disruption. The disintegrated open clusters will then become
	moving groups and supply field stars \citep{2019ApJ...877...12T}.

	The structure of the star cluster depends on various processes, such as,  star formation, gas expulsion,
        dynamics of the cluster, etc. \citep{2005ApJ...632..397G}.
	From the isodensity contours  (cf. Section 3.1), we have found that the T76 cluster is showing more or less circular morphology,
	therefore, to further quantify the structure of this cluster, we have estimate the `Q' parameter \citep[refer for details,][]{2020MNRAS.498.2309S} 
	for the sample of cluster members.
	The `Q' parameter is generally used to distinguish between clusters with a central density concentration and 
	hierarchical clusters with a fractal substructure  \citep[cf.][]{2004MNRAS.348..589C,2009MNRAS.392..341C}.
	A group of points distributed radially will
	have a high Q value (Q $>$ 0.8) while clusters with a more fractal
	distribution will have a low Q value (Q $<$ 0.8) \citep{2004MNRAS.348..589C,2014MNRAS.439.3719C}. 
	We have estimated Q=0.9 for the T76 cluster which is an
	indicative of the radial distribution of stars in this cluster.
	This is in agreement with our isodensity contour structures having circular geometry.
	As the cluster is still under the process of dynamical evolution, this radial distribution of the stars in the T76 cluster
	may be due to the star formation process itself.

	\section{Summary and Conclusion}
	\label{sect:conclusion}

	We have performed a detailed analysis of the T76 open cluster using deep NIR observations taken with the TANSPEC on the 3.6m DOT
	along with the recently available high quality PM data from the {\it Gaia} DR3 and deep photometric data from PS1.
	We have investigated the structure of this cluster, determined the membership probability of stars
	in the cluster region, derived the fundamental parameters of the cluster,
	and studied the  MF and mass segregation in this cluster.
	The main results of this study can be summarized as follows:

	\begin{itemize}

	\item
	We have have found that the T76  cluster is showing an central density concentration with circular morphology
	using the isodensity contours and the Q parameter value. This distribution is most probably due to the 
	star formation processes.
	The radius of the  T76 cluster is found to be 45$^{\prime}{^\prime}$ (1.24 pc at a distance of 5.7 kpc) centered at
         $\alpha_{2000}$: 22$^{h}$28$^{m}$46$^{s}$.68, $\delta_{J2000}$: +61$^\circ$38$^\prime$01$^\prime{^\prime}$.2.

	\item
	Using {\it Gaia} DR3 data, 28 stars were marked as highly probable cluster members.
	We have estimated the distance of this cluster both using parallax of member stars and the isochrone fitting technique
	and found that the cluster is located at a distance of $5.7\pm1.0$ kpc.
	We have also estimated the age of this cluster as $50\pm10$ Myr.

	\item
	We have derived the MF slope ($\Gamma$) in the cluster region in the mass range $\sim$0.75$<$M/M$_\odot$$<$5.8
	as $-1.3\pm0.2$ using our deep NIR data, which is similar to the  value `-1.35' given by \citet{1955ApJ...121..161S}.
	The cluster does not show any signatures of mass-segregation and is found to be
	undergoing dynamical relaxation.

\end{itemize}

\section*{Acknowledgments}

We thank the staff at the 3.6m DOT, Devasthal (ARIES) and IR astronomy group at TIFR, for their cooperation during TANSPEC observations.
This work has made use of data from the European Space Agency (ESA) mission
{\it Gaia} (\footnote{https://www.cosmos.esa.int/gaia}), processed by the {\it Gaia}
Data Processing and Analysis Consortium (DPAC,
\footnote{https://www.cosmos.esa.int/web/gaia/dpac/consortium}). Funding for the DPAC
has been provided by national institutions, in particular the institutions
participating in the {\it Gaia} Multilateral Agreement.
This publication also makes use of data from the Two Micron All Sky Survey, which is a joint project of the University of
Massachusetts and the Infrared Processing and Analysis Center/California Institute of Technology,
funded by the National Aeronautics and Space Administration and the National Science Foundation.
Part of the work/analysis was done at the National Astronomical Research Institute of Thailand.
SS acknowledged  the support of the Department of Science  and Technology,  Government of India, under project No. DST/INT/Thai/P-15/2019.
DKO acknowledged the support of the Department of Atomic Energy, Government of India, under project No. RTI 4002.


\bibliography{cz3}{}
\bibliographystyle{apj}




\end{document}